\documentclass[showpacs,prl,floatfix,preprintnumbers,twocolumn]{revtex4}

\usepackage{graphicx,url,amssymb,amsmath,longtable,rotating,color,psfig,units,wasysym,subfigure,epsfig,listings,bm}

\def\la{~\mbox{\raisebox{-.6ex}{$\stackrel{<}{\sim}$}}~}

\begin{document}
\preprint{UMN-TH-3130/12}
\preprint{LIGO-P1200172}
\preprint{IPMU12-0228}
\vspace{1cm}

\title{Measurement of Parity Violation in the Early Universe using Gravitational-wave Detectors}

\author{S. G. Crowder$^a$\footnote{crowder@physics.umn.edu}, R. Namba$^a$\footnote{namba@physics.umn.edu}, V. Mandic$^a$, S. Mukohyama$^b$, and M. Peloso$^{a,c,d}$ }

\affiliation{$^a$School of Physics and Astronomy, University of Minnesota, Minneapolis, MN 55455, USA\\
$^b$ Kavli Institute for the Physics and Mathematics of the Universe, Todai Institutes for Advanced Study, University of Tokyo, Kashiwa, Chiba 277-8583, Japan\\
$^c$Dipartimento di Fisica e Astronomia, Universit\`a di Padova\\
$^d$ INFN Sezione di Padova, I-35131 Padova, Italy
}

\date{\today}

\begin{abstract}
A stochastic gravitational-wave background (SGWB) is expected to arise from the superposition of many independent and unresolved gravitational-wave signals, of either cosmological or astrophysical origin. Some cosmological models
(characterized, for instance, by a pseudo-scalar inflaton, or by some modification of gravity) break parity, leading to a polarized  SGWB. We present a new technique to measure this parity violation, which we then apply  to the recent results from LIGO to produce the first upper limit  on parity violation in the SGWB, assuming a generic power-law SGWB spectrum across the LIGO sensitive frequency region. We also estimate sensitivity to parity violation  of the future generations of gravitational-wave detectors, both for a power-law spectrum and for a model of axion inflation. This technique   offers a new way of differentiating between the cosmological and astrophysical sources of the isotropic SGWB, as astrophysical sources are not expected to produce a polarized  SGWB.
\end{abstract}

%\pacs{95.85.Sz, 97.60.Jd, 04.25.dg, 98.80.Cq}

\bibliographystyle{unsrt_modified_2}
\maketitle

{\em Introduction.}---A stochastic gravitational-wave background (SGWB) is expected to arise from the superposition of gravitational waves (GWs) from many uncorrelated and unresolved sources. Numerous cosmological SGWB models have been proposed, including inflationary models~\cite{grishchuk,starob,Baumann:2008aq,Binetruy:2012ze}, models based on cosmic (super)strings \cite{DV1,cosmstrpaper}, and models of alternative cosmologies \cite{PBB1}. Furthermore, various astrophysical models have been proposed based on integrating contributions from astrophysical objects across the universe, such as compact binary coalescences of  binary neutron stars and/or black holes \cite{phinney,StochCBC}, magnetars \cite{cutler,marassi}, or rotating neutron stars \cite{RegPac}.
Several searches for the unpolarized isotropic \cite{S3stoch,S4stoch,S5stoch} and anisotropic SGWB \cite{S4radiometer,S5SPH} have been conducted using data acquired by interferometric GW detectors
LIGO \cite{LIGOS1,LIGOS5} and Virgo \cite{Virgo1}. These searches have established upper limits on the energy density in the SGWB, and have started to constrain some of the proposed models \cite{cosmstrpaper,StochCBC,paramest}.

In this Letter we present the first upper limits on the circularly polarized isotropic SGWB. 
Since astrophysical sources are unlikely
to produce a circularly polarized isotropic SGWB, detecting polarization
asymmetry in the SGWB is potentially an excellent way of distinguishing
the cosmological component from the possibly dominant astrophysical one.
Such asymmetry could be generated, for instance, through the gravitational Chern-Simons term \cite{alexander,setotaruya}, from the imaginary part of the Immirzi parameter \cite{Magueijo:2010ba}, in some power-counting renormalizable theories of gravity \cite{Takahashi:2009wc}, or, as we discuss in more detail, from an axion inflaton \cite{Sorbo:2011rz}.
We follow the formalism developed in \cite{allenromano}, modified to address a polarized SGWB as discussed in \cite{setotaruya}. Using the latest SGWB measurement with LIGO detectors \cite{S5stoch}, we apply this formalism to produce the first constraints on parity violation for a generic power-law SGWB spectrum.
We also estimate the sensitivity of the upcoming second-generation GW detectors to the power-law and axion-inflation parity violating models---Advanced LIGO (aLIGO) \cite{aLIGO2} detectors at Hanford, WA (H1) and Livingston, LA (L1), Advanced Virgo \cite{aVirgo} in Italy (V1), GEO-HF \cite{GEOHF} in Germany, and KAGRA \cite{CLIO,LCGT} in Japan (K1) are expected to have $\sim 10\times$ better strain sensitivities than the first-generation detectors, and to produce first science-quality data in 2015. Finally, we consider an example configuration of a pair of third-generation GW detectors, with strain sensitivity similar to the proposed Einstein Telescope \cite{ET}.

{\em Search Formalism.}---We start from the plane-wave expansion of the metric at time $t$ and position $\vec{x}$ \cite{setotaruya,allenromano}
\begin{equation}
h_{ab}(t,\vec{x}) = \sum_A\int_{-\infty}^{\infty}df \int_{S^2}d\hat{\Omega}h_A(f,\hat{\Omega})e^{-2\pi i f(t-\vec{x}\cdot\hat{\Omega})} e_{ab}^A(\hat{\Omega}),
\end{equation}
where $e_{ab}^A(\hat{\Omega})$ is the polarization tensor associated with a wave traveling in the direction $\hat{\Omega}$, and $f$ is frequency (we use  natural units $c=\hbar = 1$). We consider the  left- and right-handed correlators  \cite{setotaruya}:
\begin{eqnarray}
&&\langle h_{R/L}(f,\hat{\Omega}) h_{R/L}^*(f',\hat{\Omega}')\rangle \nonumber \\
&&\qquad = \frac{\delta(f-f') \delta^2(\hat{\Omega}-\hat{\Omega}')}{4\pi} (I(f) \pm V(f))
\end{eqnarray}
where  $h_L = (h_+ + ih_{\times})/\sqrt{2}, \; h_R = (h_+ - ih_{\times})/\sqrt{2}$, and  $+$ and $\times$ are the standard plus and cross polarizations.

\begin{figure}
\centering
\includegraphics[width=3.3in]{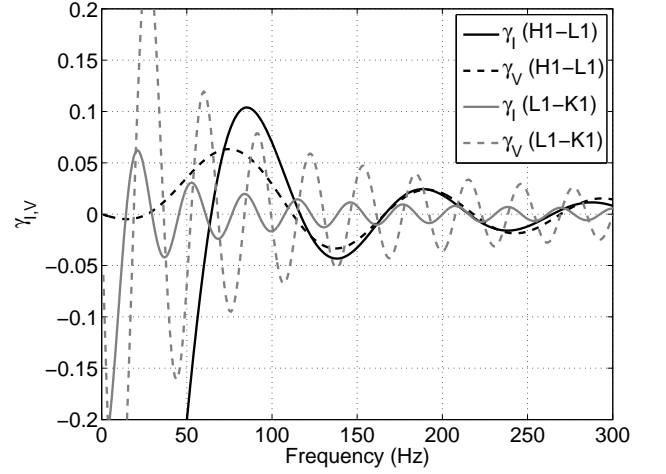}
\caption{Overlap reduction functions  for detector pairs H1-L1 and L1-K1.}
\label{gammas}
\end{figure}

Note this is the point of departure from the past searches for unpolarized isotropic SGWB, which assume $V=0$. Further note that $\langle h_R h_L^*\rangle$  vanishes due to statistical isotropy. We then compute the normalized energy density \cite{setotaruya,allenromano}:
\begin{eqnarray}
\Omega_{\rm GW}(f) = \frac{f}{\rho_c}\frac{d\rho_{\rm GW}}{df} = \frac{\pi f^3}{G_N \rho_c} I(f)
\end{eqnarray}
where $d\rho_{\rm GW}$ is the energy density in the  range $\left[ f ,\,  f+df \right]$, $G_N$ is Newton's constant, and $\rho_c$  the critical energy density of the universe \footnote{Note that the similar Eq. 3 of \cite{setotaruya} contains an additional factor of 4, which we believe is incorrect.}. We also compute the standard cross-correlation estimator \cite{allenromano}:
\begin{eqnarray}
\langle \hat{Y}\rangle & = & \int_{-\infty}^{+\infty}df \int_{-\infty}^{+\infty}df' \delta_T(f-f')\langle(s_1^*(f)s_2(f')\rangle \tilde{Q}(f') \nonumber \\
& = & \frac{3H_0^2 T}{10 \pi^2}\int_0^{\infty} df \frac{\Omega_{\rm GW}'(f) \gamma_I(f) \tilde{Q}(f)}{f^3},
\label{estimator}
\end{eqnarray}
where
\begin{eqnarray}
\label{omegaprime}
&&\Omega_{\rm GW}'(f) \gamma_I(f) = \Omega_{\rm GW}(f) \left[ \gamma_I(f) + \Pi(f) \gamma_V(f) \right] \\
&&\gamma_I(f) = \frac{5}{8\pi} \int d\hat{\Omega} (F_1^+ F_2^{+*} + F_1^{\times} F_2^{\times *}) e^{2\pi i f \hat{\Omega} \cdot \Delta \vec{x}} \nonumber \\
&&\gamma_V(f) = -\frac{5}{8\pi} \int d\hat{\Omega} i(F_1^+ F_2^{\times*} - F_1^{\times} F_2^{+*}) e^{2\pi i f \hat{\Omega} \cdot \Delta \vec{x}}. \nonumber
\end{eqnarray}
Here, $T$ is the measurement time, $\delta_T(f)\equiv\sin(\pi fT)/(\pi f)$, $\tilde{s}_1(f)$ and $\tilde{s}_2(f)$ are Fourier transforms of the strain time-series of two GW detectors, $\tilde{Q}(f)$ is a filter, and $F_n^A = e_{ab}^A d^{ab}_n$ is the contraction of the tensor mode of polarization $A$, $e_{ab}^A$, with the response of the detector $n$, $d^{ab}_n$ \footnote{Note that the minus sign in the Equation for $\gamma_V(f)$ is missing in the  computation of \cite{setotaruya}.}. The factor $\gamma_I(f)$ is the standard overlap reduction function arising from different locations and orientations of the two detectors, and $\gamma_V(f)$ is a new
function, associated with the parity violating term. Figure \ref{gammas} shows these functions for two real detector pairs. Finally, $\Pi(f)=V(f)/I(f)$ encodes the parity violation, with maximal values $\Pi=\pm 1$ corresponding to fully right- or left-handed polarizations.   Setting $\Pi=0$ reproduces the standard unpolarized SGWB search \cite{allenromano}.

Assuming stationary Gaussian detector noise (uncorrelated between two detectors), the estimator for the variance associated with $\hat{Y}$ is  \cite{allenromano}:
\begin{equation}
  \sigma^2 = \frac{T}{4} \int_{0}^{\infty} df P_1(f) P_2(f) |\tilde{Q}(f)|^2\,,
  \label{sigma}
\end{equation}
where $P_n(f)$ are the one-sided noise power spectral densities of the two GW detectors.
In practice, we divide the sensitive frequency band of the GW detectors into bins $\Delta f=0.25$ Hz wide \cite{S5stoch}. We then compute the estimator $\hat{Y}_i$ and the variance $\sigma_i^2$ for each frequency bin $i$ assuming 
a frequency independent spectrum template $\Omega_{\rm GW}(f)=\Omega_0$ for each bin. Optimization of the signal-to-noise ratio then leads to the following optimal filter for a frequency-independent GW spectrum in the frequency bin $f_i$~\cite{allenromano}:
\begin{equation}
  \label{optfilt1}
  \tilde{Q}(f_i) = \mathcal{N} \; \frac{\gamma_I(f_i)}{f_i^3 P_1(f_i) P_2(f_i)} \; ,
\end{equation}
with normalization constant $\mathcal{N}$  chosen so that  $\langle \hat{Y}_i\rangle = \Omega_0$.

To perform parameter estimation in the parity violating models, we adopt the Bayesian approach introduced in \cite{paramest}. We define the following likelihood function:
\begin{equation}
  L(\hat Y_i, \sigma_i|\vec\theta) \propto \exp \left[ -\frac{1}{2}
    \sum_i \frac{(\hat Y_i - \Omega_{\rm M}'(f_i; \vec\theta))^2}{\sigma_i^2} \right]
\label{sgwblik}
\end{equation}
where the sum runs over frequency bins $f_i$, and $\Omega_{\rm M}'(f_i; \vec\theta)$ is the SGWB energy density spectrum specified by the free parameters $\vec{\theta}$, multiplied by the parity-violating correction $(1+\Pi(f) \gamma_V(f)/\gamma_I(f))$, as defined in Eq. \ref{omegaprime}. Multiplying the likelihood with the prior distribution for $\vec\theta$ yields the Bayesian posterior distribution, which can then be used to extract confidence intervals for $\vec\theta$. In the subsequent results, we take all priors to be flat within the plotted range of $\vec\theta$ (and zero elsewhere).

{\em Power Law.}---We  apply this  formalism to the power-law spectrum, $\Omega_{\rm M}(f) = \Omega_{\alpha} (f/f_{ref})^{\alpha}$, which is appropriate for most SGWB models in the LIGO frequency band. We set $f_{ref}=100$ Hz, and  assume $\Pi(f) = \Pi = const$.
\begin{figure*}[!t]
   \begin{tabular}{cc}
    \psfig{file=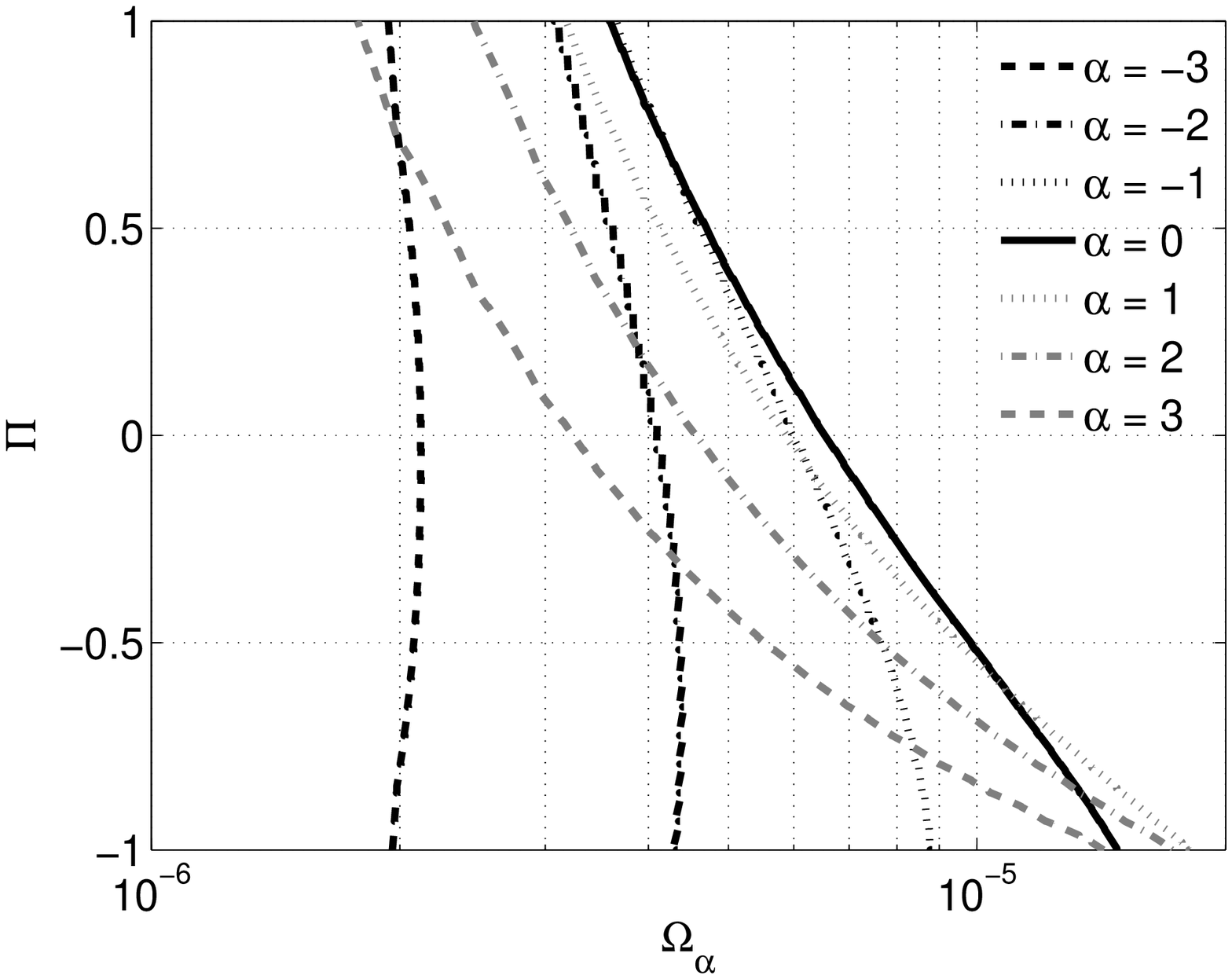, height=2.5in} &
    \psfig{file=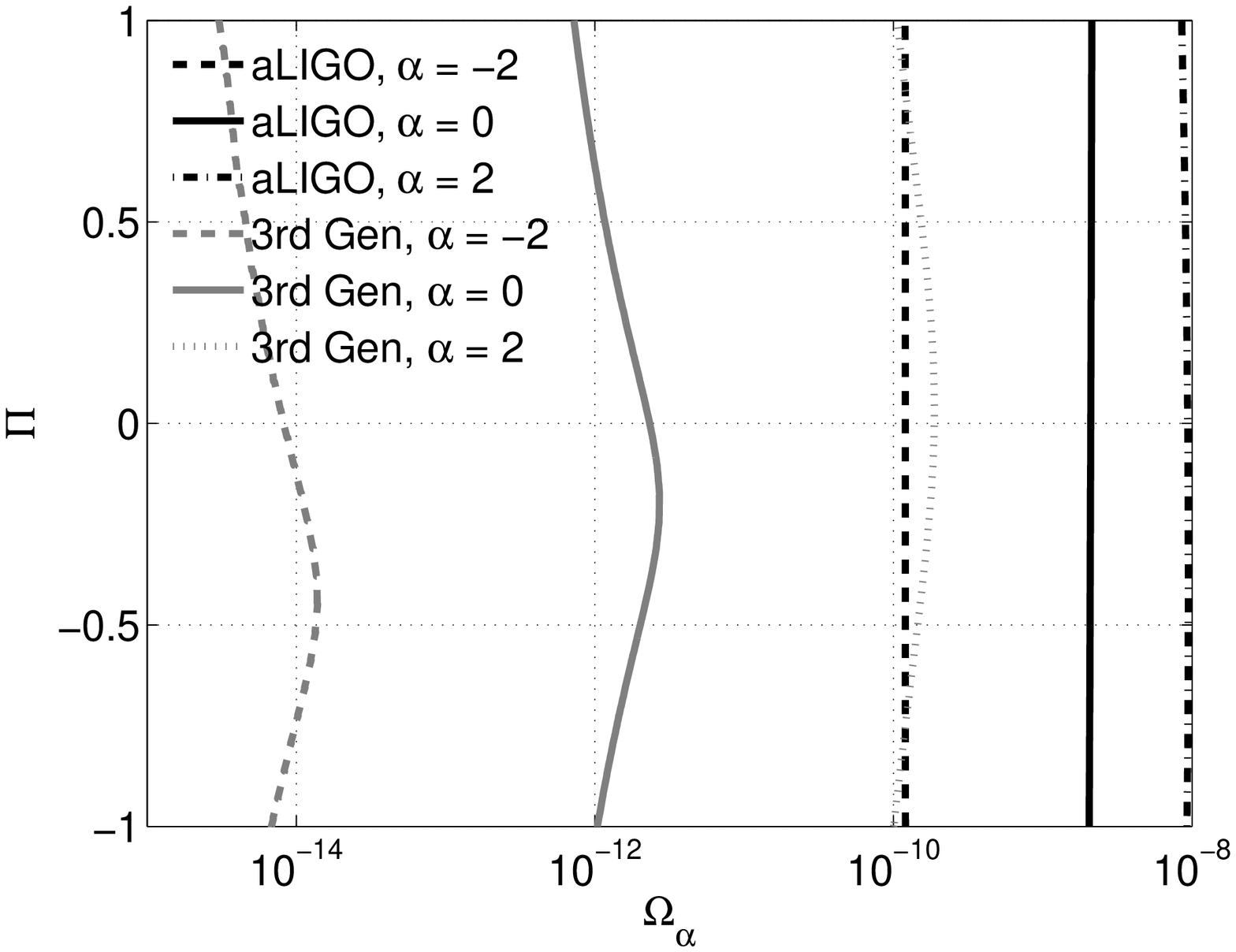, height=2.5in} \\
    \psfig{file=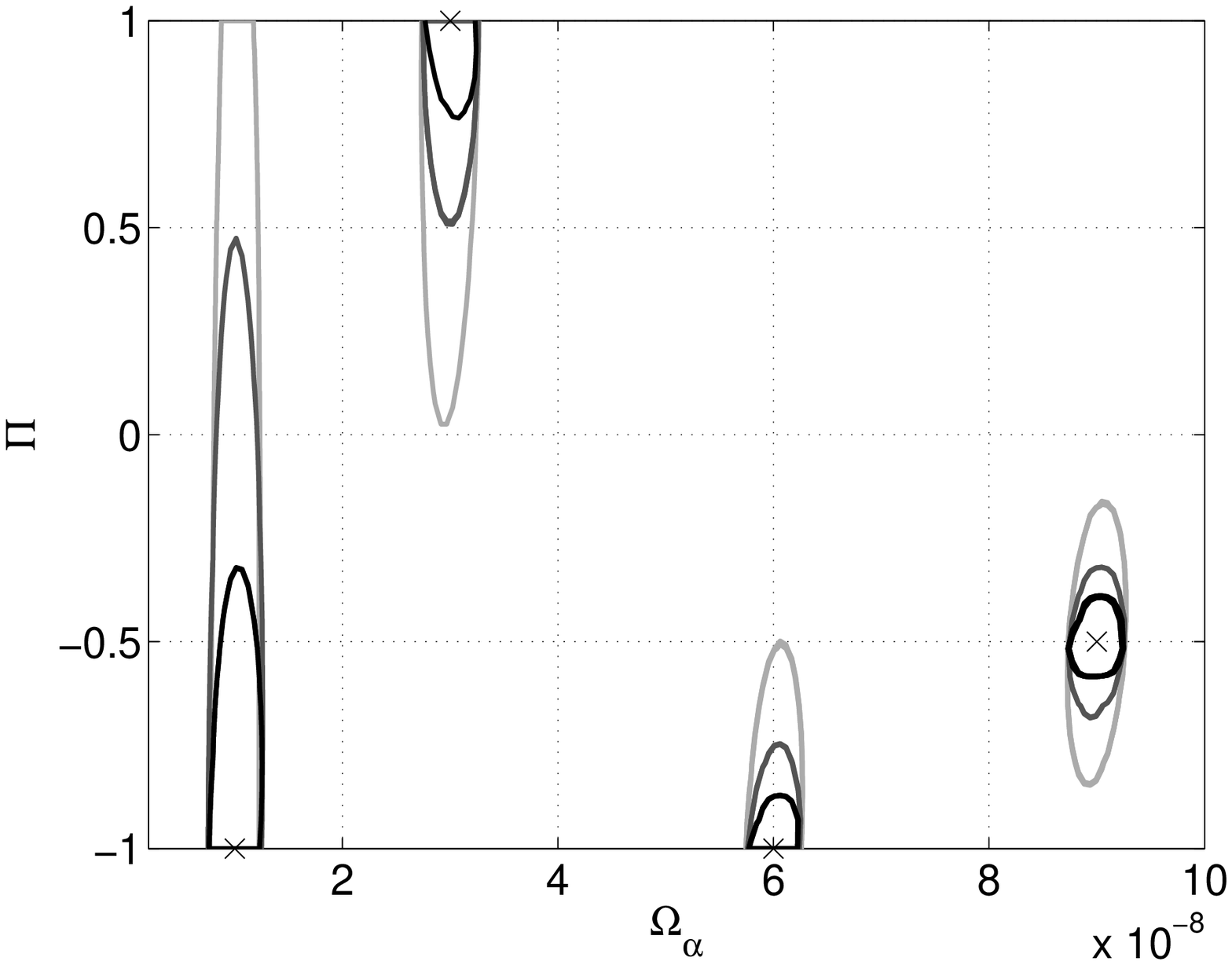, height=2.5in} &
    \psfig{file=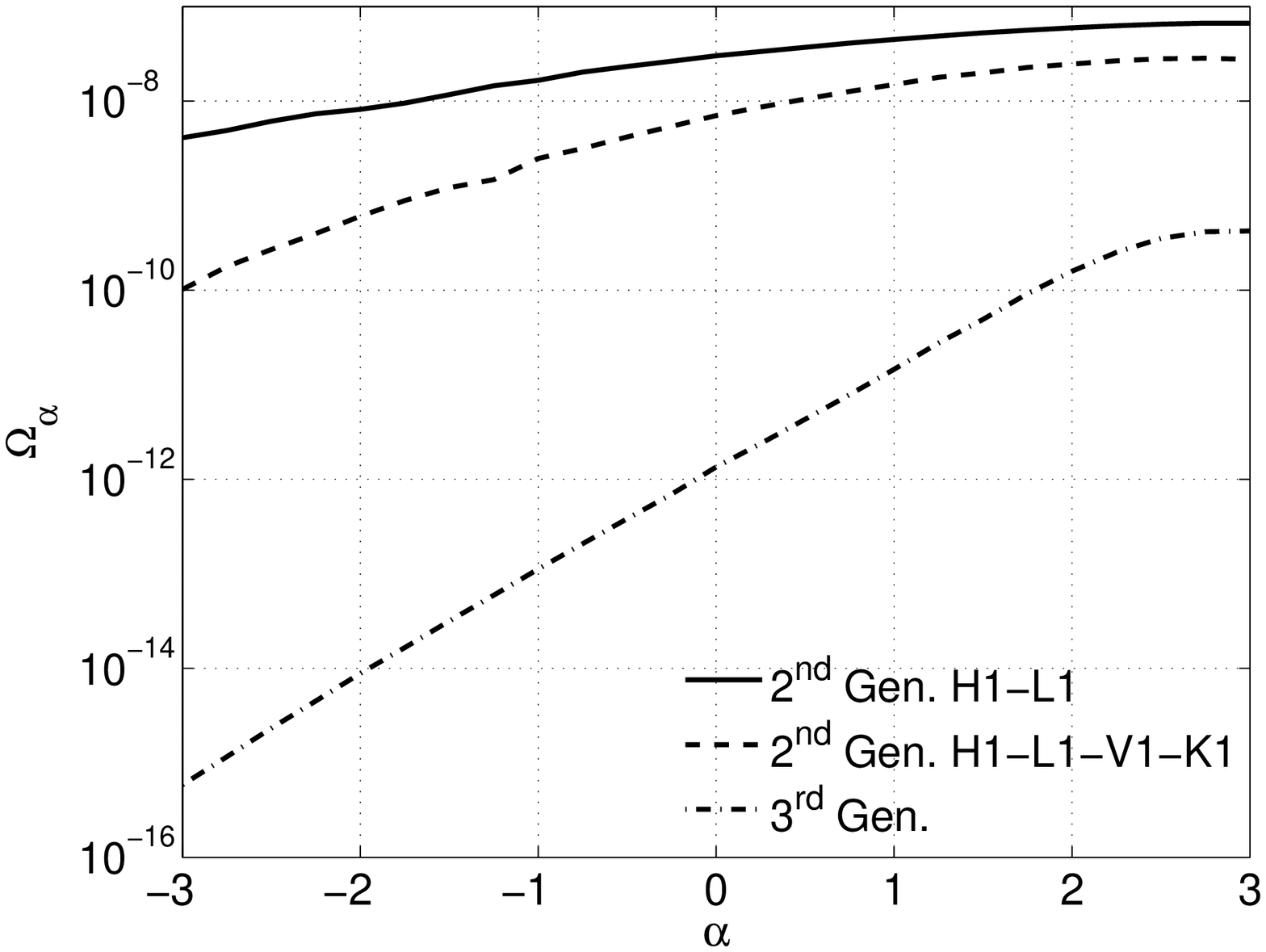, height=2.5in}
  \end{tabular}
\caption{Top-left: 95\% CL limit curves based on the most recent LIGO result \cite{S5stoch}, for several values of $\alpha$. Top-right: Expected sensitivities (at $2\sigma$ level) for the advanced LIGO H1-L1 pair and for an example of a third-generation detector pair (see text for more detail). Bottom-left: 95\% CL contours are shown for several examples of simulated SGWB with parity violation, assuming $\alpha=0$, standard strain sensitivities, and 1 year of observation. The x's denote the signal simulation parameter values. The lightest-gray line corresponds to the recovery with H1-L1, medium-gray line to the recovery with H1-L1-V1, and the black line to the recovery with H1-L1-V1-K1 network.
Bottom-right: assuming a SGWB with maximal parity violation ($\Pi=+ 1$), the lines denote $\Omega_{\alpha}$ needed for a given $\alpha$ to detect the SGWB and exclude $\Pi=0$ at 95\% confidence, using two second-generation detector networks and the example of a third-generation detector pair.}
\label{powerlawcontours}
\end{figure*}

The results are shown in Figure \ref{powerlawcontours}. The top-left plot of the Figure shows the 95\% CL exclusion curves in the $\Pi-\Omega_{\alpha}$ plane for several values of the spectral index $\alpha$, using the latest SGWB measurement with LIGO detectors (specifically, using the data plotted in Figure 6 of \cite{S5stoch}). This plot represents the first constraint on parity violation in the early universe using GW observations. The top-right plot of Figure \ref{powerlawcontours} shows the expected sensitivities for the detection of a SGWB at
a $2 \sigma$ level of the Advanced LIGO H1-L1 detector pair, assuming
standard expected strain sensitivity \cite{aLIGO2}. It also shows the sensitivity of an example
configuration of two third-generation detectors, assuming two L-shape
detectors of strain sensitivity similar to the Einstein Telescope design
\cite{ET}, located at opposite poles of the Earth, with respective arm
orientation rotated by $\pi/8$. Such a pair of detectors has not been proposed yet, so this example is only meant to illustrate the potential reach of third-generation detectors, in a configuration that suppresses neither $\gamma_I$ nor $\gamma_V$ terms.
In both cases we assume 1 year of observation. As expected, the second-generation detectors will provide 3-4 orders of magnitude better SGWB measurement as compared to the first-generation detectors, 
and the third-generation detectors could provide yet another 3-4 orders of magnitude improvement.

The bottom-left plot of Figure \ref{powerlawcontours} illustrates the recovery of several simulated parity violating SGWB signals using different networks of second-generation GW detectors. Having multiple detector pairs in the network does not significantly improve the theoretical uncertainty $\sigma_i$ (Eq. \ref{sigma}) over the H1-L1 detector pair. However, multiple detector pairs introduce multiple $\gamma_V$'s in the search, which helps break the degeneracy between the parity conserving and violating
terms in Eq. \ref{omegaprime}. As a result, the 95\% CL contours are significantly tighter in $\Pi$ when using multiple detector pairs. Finally, the bottom-right plot shows the amplitude $\Omega_{\alpha}$ required to detect parity violation with the H1-L1 and H1-L1-V1-K1 second-generation detector networks, and with the third-generation example pair (assuming maximal parity violation in the SGWB, $\Pi=+1$). In particular, for $\alpha=0$ the H1-L1-V1-K1 detector network could detect parity violation at amplitudes $4\times$ smaller than the H1-L1 pair alone, while the third-generation detectors could detect it at amplitudes another $\sim 4$ orders of magnitude smaller.

{\em Parity violation during inflation.}---We now apply the above formalism to a specific example of a parity violating SGWB. We concentrate on an example characterized by standard gravity. Specifically,  we consider a natural vast class of inflationary models where the flatness of the potential $V$ of the inflaton $\phi$ is protected by  an approximate shift symmetry $\phi \rightarrow \phi + {\rm constant}$. The most studied realization of this is the axion field \cite{Peccei:1977hh}, and many models, starting from  \cite{Freese:1990rb}, have employed this idea in the context of inflation. Generally, after inflation an axion inflaton predominantly decays into gauge quanta through the coupling $\Delta {\cal  L} \sim \frac{C}{f_{\phi}} \phi F  {\tilde F}$, where  $f_{\phi}$ is the axion decay constant, $C$ is a parameter which is naturally of order one, $F$ is the field strength of the gauge field (for simplicity, a U(1) field is considered here), and ${\tilde F}$ its dual.   This interaction leads to an interesting gauge field production also during inflation \cite{Anber:2009ua}, as it modifies the dispersion relation of the two gauge field helicities to
\begin{equation}
\omega_\pm^2 = k^2 \mp 2 k a H \xi \;\;\;,\;\;\; \xi \equiv \frac{C \, \dot{\phi}}{2 f_{\phi} H} \;\;.
\end{equation}
Here, $a$ is the scale factor of the universe that grows nearly exponentially during inflation,
$a \simeq {\rm e}^{H t}$, while the other dynamical quantities are  slowly varying. Any gauge field mode is identified by its constant comoving momentum $k$. The second term in $\omega_\pm^2$ becomes dominant as the  Compton wavelength $\lambda \sim a/k$ becomes greater than the  Hubble horizon $H^{-1}$ (this moment is dubbed ``horizon crossing'';  we denote by $\xi_k$ the value of $\xi $  at this moment). As a consequence,
one  of the two helicities experiences a tachyonic growth.  For definiteness, we  assume $\xi > 0$, so that the growing helicity is  the $+$ one.   Modes of progressively smaller $\lambda$ (larger $k$) cross the horizon later during inflation. Since $\xi \propto \frac{\dot{\phi}}{H}$ grows during inflation, $\xi_k$ is a slowly increasing function of $k$. Since the amplitude of the produced gauge modes is $\propto {\rm e}^{\pi \xi_k}$ \cite{Anber:2009ua}, even this slow increase can result in a much stronger production at short wavelengths.

The produced gauge quanta generate inflaton perturbations and GW through $A_+ A_+ \rightarrow \delta \phi$ and  $A_+ A_+ \rightarrow h$ processes \cite{Barnaby:2010vf}. These ``sourced'' signals superimpose incoherently to the standard ``vacuum'' inflaton and metric perturbations. 
For example, the energy density in GW generated during inflation is $\Omega_{\rm GW}  \propto   \langle h_{\rm vacuum}^2 \rangle  +   \langle h_{\rm sourced}^2 \rangle $, and identically for $\delta \phi$. The  $A_+ A_+ \rightarrow h$ interaction mostly produces GW of left chirality \cite{Sorbo:2011rz}, and therefore $\Pi \simeq - 1$ whenever $h_{\rm sourced} \gg h_{\rm vacuum}$ (equivalently, $\Pi \simeq + 1$ for $\xi<0$). The $\delta \phi_{\rm sourced}$ signal is highly non-gaussian \cite{Barnaby:2010vf,Barnaby:2011vw}, while the primordial  density perturbations are highly gaussian. Therefore,  $\delta \phi_{\rm sourced}$ needs to be subdominant to the vacuum signal, which, according to the estimate of  \cite{Barnaby:2011vw},  results in the constraint  $\xi_{\rm CMB} \la 2.66$ at the moment in which the modes of cosmological scales (those that determine the Cosmic Microwave Background anisotropies) were produced~\footnote{This corresponds to about $60$ e-folds before the end of inflation. See   \cite{Barnaby:2011qe} for the precise definition. Our limits and forecasts for $\xi$ refer to the value assumed at this moment.}.
When this limit is respected, $h_{\rm sourced} \ll h_{\rm vacuum}$ on cosmological scales, and is unobservable
\cite{Barnaby:2010vf,Sorbo:2011rz,Barnaby:2011vw} in the minimal implementations of the mechanism. However, given the increase of $\xi_k$ with $k$,  the produced $A_+$  and, consequently,  the amount of $h_{\rm sourced}$ steeply  increases
with increasing frequency \cite{Cook:2011hg}, to a level that can be observed by terrestrial GW detectors. The  computation of this effect requires taking into account the backreaction of the produced gauge fields on the late inflaton dynamics \cite{Barnaby:2011qe}, and leads to the results summarized in Figures 5 and 6 of that work. It was shown there that the signal can be observed by Advanced LIGO for values of $\xi$ that do not conflict with the CMB bound.

We apply the formalism developed above to this inflation model. In particular, we consider two power-law inflaton potentials, $V \sim \phi^p$ for $p = 1$ (typical of axion monodromy \cite{McAllister:2008hb}) and $p=2$ (a generic Taylor expansion around the minimum of a potential).   For each $p$,   $\Omega_{\rm M} (f; \xi)$ in Eq. \ref{sgwblik} is a function of the single parameter $\xi$.
We then compute the expected $2\sigma$ sensitivities to $\xi$ for the second-generation H1-L1-V1-K1 detector network and for the example third-generation detector pair, assuming flat prior distribution for $\xi$ up to the CMB bound of \cite{Barnaby:2011vw}.  The results are shown in Table \ref{table1}. The second-generation detectors will begin to constrain the $\xi$ parameter beyond the CMB gaussianity constraint of \cite{Barnaby:2011vw}, while the third-generation detectors may reach $\xi \sim 1.8$. Somewhat larger values of $\xi$ are required in order to detect parity violation in the observed SGWB and rule out $\Pi=0$ at the $2\sigma$ level.
\begin{table}
%\centering
\begin{tabular}{|r|r|r|r|r|r|}
\hline
Detector Network & p & $\xi$ (sensitivity) & $\xi$ (exclude $\Pi=0$)\\
\hline
$2^{\rm nd}$ Gen H1-L1-V1-K1 & \;\;1 & 2.3 & $> 2.66$ \\
$2^{\rm nd}$ Gen H1-L1-V1-K1 & \;\;2 & 2.2 & 2.6 \\
$3^{\rm rd}$ Gen & \;\;1 & 1.8 & 2.0 \\
$3^{\rm rd}$ Gen & \;\;2 & 1.9 & 2.0 \\
\hline
\end{tabular}
\caption{Column 3 shows the $2\sigma$ sensitivity to $\xi$ for different future GW networks (column 1) in the axion-based inflation model discussed in the text, for two power-law inflaton potentials: $p=1$ or $p=2$ (column 2). Column 4 indicates the value of $\xi$ needed in order for the respective detector network to detect parity violation (reject $\Pi=0$) at $2\sigma$.}
\label{table1}
\end{table}

We note that $\xi$ can also be constrained by taking advantage of its impact on $\delta \phi_{\rm sourced}$, which leads to the non-gaussian signature mentioned above, to the growth of the power spectrum of the scalar perturbations at the highest values of $k$ that can be observed in the CMB \cite{Meerburg:2012id}, and to the production of primordial black holes \cite{Linde:2012bt}.
The bounds from the first two effects  \cite{Meerburg:2012id}  are weaker than those from the SGWB. Those from the primordial black holes are potentially stronger, but they carry an uncertainty associated with how many scalar perturbations are produced at such small scales, and with how many black holes they create  \cite{Linde:2012bt}. No such uncertainties are present for the SGWB signal that we have studied here.

{\em Conclusions.}---We introduce a new formalism for measuring polarization asymmetry in the SGWB, which both improves the sensitivity of the SGWB
searches and offers a new handle for distinguishing between the cosmological and astrophysical SGWB sources. We apply the formalism to the recent LIGO data to produce the first constraints on the parity violation in the SGWB. We also estimate the sensitivities of future GW detectors to the parity violating SGWB signals in the framework of a generic power-law SGWB spectrum and in the specific example of an axion-based inflation model, demonstrating the substantial advantage of networks allowing multiple non-collocated detector pairs.

\smallskip {\bf Acknowledgments}---The work of SGC and VM was in part supported by the NSF grant PHY1204944. The work of RN and MP was supported in part  by DOE grant DE-FG02-94ER-40823 at the University of Minnesota. The work of SM was supported by Grant-in-Aid for Scientific Research 24540256 and 21111006, by Japan-Russia Research Cooperative Program and by the WPI Initiative, MEXT, Japan. MP thanks the University and the INFN of Padua for their friendly hospitality and for partial support during his sabbatical leave. The authors thank LIGO and Virgo Collaborations for providing the data plotted in Figure 6 of \cite{S5stoch}.

%\nocite{*}
\bibliography{parviol}

\end{document}